\documentclass[sigplan,screen,nonacm]{acmart}

\AtBeginDocument{%
  }

\setcopyright{none}
\renewcommand\footnotetextcopyrightpermission[1]{}
\usepackage{listings}
\usepackage{xcolor}
\usepackage{makecell}
\usepackage{multirow}

\usepackage[utf8]{inputenc}

\usepackage{listings}
\usepackage{xcolor}

\lstdefinelanguage{JavaScript}{
  sensitive=true,
  morecomment=[l]{//},
  morecomment=[s]{/*}{*/},
  morestring=[b]',
  morestring=[b]",
  morekeywords={typeof,new,true,false,catch,function,return,null,break,continue,
                for,while,if,else,in,instanceof,let,const,var,of,try,throw,
                case,switch,default,this},
}

\lstdefinestyle{jsblock}{
  language=JavaScript,
  basicstyle=\ttfamily\footnotesize,
  showstringspaces=false,
  keepspaces=true,
  columns=fullflexible,
  breaklines=true,
  breakatwhitespace=true,
  frame=single,
  rulecolor=\color{black!30},
  backgroundcolor=\color{black!1},
  xleftmargin=0pt, aboveskip=4pt, belowskip=4pt
}

\begin{document}

\title{UI-CUBE: Enterprise-Grade Computer Use Agent Benchmarking Beyond Task Accuracy to Operational Reliability}

\author{Horia Cristescu}
\email{horia.cristescu@uipath.com}
\affiliation{%
  \institution{UiPath}
  \country{Romania}
}

\author{Charles Park}
\email{charles.park@uipath.com}
\affiliation{%
  \institution{UiPath}
  \country{UK}
}

\author{Trong Canh Nguyen}
\email{canh.nguyen@uipath.com}
\affiliation{%
  \institution{UiPath}
  \country{France}
}

\author{Sergiu Talmacel}
\email{sergiu.talmacel@uipath.com}
\affiliation{%
  \institution{UiPath}
  \country{Romania}
}

\author{Alexandru-Gabriel Ilie}
\email{alexandru.ilie@uipath.com}
\affiliation{%
  \institution{UiPath}
  \country{Romania}
}

\author{Stefan Adam}
\email{stefan.adam@uipath.com}
\affiliation{%
  \institution{UiPath}
  \country{Romania}
}

\renewcommand{\shortauthors}{Cristescu et al.}

\begin{abstract}
While current Computer Use Agent (CUA) benchmarks measure task completion effectively, they provide limited assessment of enterprise deployment readiness, emphasizing functional correctness over the operational reliability required for production systems. We present UI-CUBE (UiPath Computer Use BEnchmark), a systematic benchmark comprising 226 tasks across two difficulty tiers designed to expose fundamental architectural limitations in current CUAs. Our evaluation covers simple UI interactions (136 tasks) and complex workflows including copy-paste tasks (50 tasks) and enterprise application scenarios (40 tasks), with systematic interface variation coverage, multi-resolution testing and automated validation of task success through the application state.

Evaluation of five state-of-the-art models reveals a sharp capability cliff rather than gradual performance degradation. Simple UI interactions achieve 67-85\% success rates (compared to 97.9\% human performance), but complex workflows drop precipitously to 9-19\%. Human evaluators with no prior application experience achieve only 61.2\% on complex tasks despite near-perfect performance on simple tasks, establishing realistic performance ceilings. This discontinuous performance pattern---where agents achieve 68-87\% of human performance on simple tasks but only 15-32\% on complex workflows---indicates fundamental architectural limitations in memory management, hierarchical planning, and state coordination rather than incremental capability gaps addressable through better training or prompting.

UI-CUBE functions as an enterprise-readiness diagnostic, revealing that while current CUAs can manipulate individual interface elements, they cannot yet function as reliable workflow automation tools. These findings provide architectural insights essential for developing production-ready CUAs capable of managing complex, multi-step enterprise processes.
\end{abstract}

%% \acmConference{Preprint}{2025}{}

\maketitle

%%
%% Repository link
\noindent\textbf{GitHub:} {\small\url{https://github.com/UiPath/uipath_enterprise_benchmark}}

\section{Introduction}

Computer Use Agents (CUAs) are rapidly emerging as a new way to automate digital tasks across diverse software environments. Despite their promise, existing benchmarks primarily measure functional task completion without systematically assessing the consistency, robustness, and reliability required for enterprise deployment. Questions remain about CUA capabilities across the full spectrum of interface controls and interaction patterns, their performance stability across different screen resolutions, and their ability to coordinate multi-step workflows with the operational reliability demanded by production systems.

To close this gap, we designed UI-CUBE to provide two complementary assessments: systematic coverage of atomic UI interactions to map interface-level capabilities, and faithful mocks of complex enterprise workflows to test coordination and operational reliability. This includes testing across screen resolutions, validating performance with workflows from SAP, Workday, and other enterprise systems, and evaluating multi-step processes where reliability is as critical as task completion.

With this in mind, we designed UI-CUBE, a new benchmark tailored to enterprise-like conditions, and used it to evaluate our own CUA implementation alongside state-of-the-art systems. By testing against these realistic scenarios, we were able to measure not only functional accuracy but also operational reliability---providing clearer insight into how CUAs can move from experimental demos toward dependable enterprise-grade tools.

\section{Related Work}

The emergence of Computer Use Agents has driven development of increasingly sophisticated evaluation frameworks, each revealing different aspects of current capabilities and limitations.

\textbf{Early Web UI Benchmarks.} The foundation for web-based agent evaluation was established by MiniWoB and MiniWoB++~\cite{shi2017world,liu2018reinforcement}, which provided simplified web interaction tasks for reinforcement learning agents. While these environments demonstrated the feasibility of programmatic UI evaluation, their synthetic, template-based tasks represented controlled conditions rather than the complexity of real enterprise applications. Mind2Web~\cite{deng2023mind2web} advanced toward realism by collecting diverse tasks from real-world websites with comprehensive webpage snapshots for offline replay, though its reliance on non-executable captures rather than containerized environments limits agents to deterministic trace replay rather than interactive exploration.

\textbf{OSWorld}~\cite{xie2024osworld} establishes the gold standard for desktop environment evaluation with 369 real tasks across Ubuntu, Windows, and macOS, using virtual machines and execution-based validation scripts. Their findings reveal fundamental gaps: human performance reaches 72.36\% while agents at initial publication achieved approximately 12\%. Recent systems on the OSWorld leaderboard have improved to approximately 60\% (Agent S3: 62.6\% single-run, 69.9\% with Best-of-10), though performance varies significantly with max steps allowed (50-100 vs unlimited) and whether coding-based actions are enabled. A separate efficiency study~\cite{abhyankar2025osworld} found agents require 1.4-2.7$\times$ more steps than humans for identical tasks. OSWorld's systematic analysis of GUI grounding failures---mis-clicks, focus loss, and dialog handling errors---provides crucial diagnostic insights. Platform-specific benchmarks complement this multi-OS evaluation: Windows Agent Arena~\cite{bonatti2024windows} focuses on the dominant desktop OS with 150+ tasks achieving 19.5\% agent versus 74.5\% human success, AndroidWorld~\cite{rawles2024androidworld} addresses mobile platforms with 116 programmatic tasks reaching 30.6\% success, and macOSWorld~\cite{yang2025macosworld} introduces multilingual evaluation across 202 tasks revealing significant cross-language performance degradation. These platform-specific efforts collectively validate that resolution and environment variability represent fundamental rather than peripheral evaluation dimensions. Our work builds on OSWorld's environmental sophistication but shifts from broad task coverage to systematic interface taxonomy, starting from identified failure modes and systematically expanding to map the space of enterprise UI interactions rather than organic task sampling to expose architectural capability boundaries.

\textbf{WebArena}~\cite{zhou2023webarena} pioneered self-hosted web environments with realistic multi-site scenarios spanning e-commerce, forums, and content management. However, their evaluation methodology reveals critical validation weaknesses: LLM judges accept incorrect trajectories (marking ``45 + 8 minutes'' as correct for ``63 minutes'') and brittle string matching penalizes valid alternative solutions. These failures motivated our programmatic postcondition validation that checks final application state rather than model-as-oracle approaches. REAL~\cite{garg2025real} advances deterministic web evaluation through 11 high-fidelity website replicas with 112 tasks, providing hosted environments and programmatic state verification similar to our approach. Their findings that frontier models achieve at most 41\% success (Claude 3.7-Sonnet) reinforce the challenge of reliable web automation. Where WebArena and REAL focus on web-specific interactions, our benchmark addresses the broader enterprise application landscape through mocked business systems.

\textbf{SWE-bench}~\cite{jimenez2023swe} transforms real GitHub issues into unit-testable agent tasks, providing authentic software engineering challenges. Subsequent audits revealed that incomplete test suites can inflate performance estimates by up to 100\% and reorder entire leaderboards, demonstrating that deterministic evaluation fails without comprehensive oracle quality. This finding informs our validation approach, where all tasks use the same programmatic test() function methodology but validate increasingly complex goal states---from simple interface changes to multi-step workflow completion to enterprise process outcomes. Our diagnostic framing extends SWE-bench's insight that evaluation methodology directly impacts capability assessment. Complementing evaluation work, Agent Workflow Memory~\cite{wang2024agentmemory} demonstrates that architectural innovations—inducing reusable task workflows from trajectories—can substantially improve performance (24.6\% on Mind2Web, 51.1\% on WebArena), suggesting that addressing the memory management limitations we identify represents a promising research direction.

\textbf{TheAgentCompany}~\cite{xu2024theagentcompany} simulates a complete software company environment, measuring cost, collaboration quality, and trajectory analysis across development workflows. While advancing enterprise-focused evaluation, it remains bounded to a single organizational context. OfficeBench~\cite{wang2024officebench} evaluates multi-application office automation across Word, Excel, PDF, Calendar, and Email with 300 tasks, finding GPT-4 Omni achieves only 47\% success despite relatively constrained task scopes. Their observation of performance degradation from single-app (64.52\%) to three-app scenarios (21.43\%) validates our finding that application coordination represents a fundamental challenge. Our enterprise application tier provides targeted coverage of ERP, CRM, and HR systems through faithful mocks of key workflows, offering broader assessment of low-level business software interaction capabilities that complement TheAgentCompany's deep organizational integration and OfficeBench's cross-application evaluation.

\textbf{WebVoyager}~\cite{he2024webvoyager} targets live websites with multimodal agents using GPT-4V~\cite{openai2023gpt4} as an automatic judge, claiming high human agreement in evaluation. However, their reliance on LLM evaluation inherits the validation fragility issues identified in WebArena, where judge quality determines benchmark reliability. Our systematic interface coverage approach addresses this by providing structured evaluation criteria based on empirical interaction patterns rather than depending on model judgment consistency.

Concurrent work SCUBA~\cite{dai2025scuba} provides complementary depth on enterprise CRM workflows, evaluating 300 Salesforce-specific tasks with fine-grained milestone-based metrics. Where UI-CUBE provides systematic coverage across enterprise application patterns (CRM, ERP, HR, expense management), SCUBA offers deep task diversity within a single platform. CRMArena-Pro~\cite{huang2025crmarenapro} extends CRM evaluation with 19 tasks across Sales, Service, and CPQ scenarios, incorporating multi-turn interactions and confidentiality awareness assessment. Critically, CRMArena-Pro observes similar capability degradation: leading agents achieve approximately 58\% success in single-turn scenarios but drop to 35\% in multi-turn settings, validating that workflow coordination represents a cross-benchmark challenge. All three benchmarks validate task success through programmatic inspection of application state rather than trajectory analysis or LLM judges, addressing validation brittleness identified in prior work.

\textbf{WorkArena++}~\cite{boisvert2024workarena} provides critical validation of the capability cliff phenomenon through 682 enterprise workflow tasks on ServiceNow, structured across three difficulty tiers: L1 (atomic tasks like form filling), L2 (compositional workflows with explicit step-by-step instructions), and L3 (realistic workflows via ticket assignments requiring knowledge base consultation). Their results reveal the same discontinuous performance pattern we observe: GPT-4o achieves 42.7\% on atomic tasks but drops to 3\% on compositional workflows and 0\% on complex reasoning tasks, while humans maintain 93.9\% success. This cross-benchmark validation strengthens the conclusion that current performance degradation stems from fundamental architectural constraints rather than task-specific evaluation artifacts. Where WorkArena++ focuses on compositional planning within a single enterprise platform, UI-CUBE provides systematic coverage across interface diversity and application types, enabling complementary insights into agent limitations.

These benchmarks collectively reveal that current evaluation approaches suffer from outcome myopia---measuring task completion without assessing the consistency and robustness across interface variations, resolutions, and workflow complexity required for enterprise deployment. While the field has progressed from template-based environments~\cite{shi2017world,liu2018reinforcement} to realistic web scenarios~\cite{zhou2023webarena} and desktop systems~\cite{xie2024osworld}, evaluation remains focused on organic task sampling rather than systematic interface coverage. Our two-tier diagnostic design addresses this gap by exposing the sharp capability cliffs between simple UI manipulation and coordinated workflow execution, revealing specific architectural limitations rather than only ranking aggregate performance. Where existing benchmarks optimize for comprehensive task coverage or realistic environments, our evaluation reveals that while agents handle diverse interface patterns reasonably well (67-85\%), they fail catastrophically on coordinated workflows (9-19\%), indicating fundamental limitations in memory management and multi-step coordination that prevent enterprise adoption.

\section{Benchmark Design}

UI-CUBE evaluates Computer Use Agents in enterprise environments. Our design targets two primary failure modes observed in current systems: interface adaptation across UI variations and workflow coordination in multi-step processes. All tasks are evaluated across multiple screen resolutions to assess agent consistency under different display conditions.

\subsection{Task Design and Empirical Coverage}

Our 226 tasks span two difficulty tiers: simple UI interactions (136 tasks) and complex workflows (90 tasks comprising 50 copy-paste/business-process tasks and 40 enterprise application tasks). After implementation, we classified tasks along three dimensions to measure empirical coverage: 22 control types, 27 structure types, and 27 action types. Action distributions reflect real-world automation patterns with selection operations most frequent (23\%), followed by navigation (16.1\%) and typing (11.1\%). This mirrors enterprise workflows where agents primarily manipulate existing interface elements rather than generate content.

\subsection{Simple Scenarios: Maximizing Interface Diversity}

\begin{table*}
  \caption{Simple Scenario Task Distribution}
  \label{tab:simple-scenarios}
  \begin{tabular}{p{3cm}p{4cm}cp{4.5cm}}
    \toprule
    Control Category & Interaction Patterns & Count & Key Variations\\
    \midrule
    Combo Box & Autocomplete, browse-heavy, token-based & 20 & Fuzzy matching, async validation, multi-selection \\
    Date Picker & Calendar pickers, range selection & 20 & Modal vs inline, multi-date, timezone handling \\
    Time Picker & Clock faces, dropdowns, sliders & 20 & 12/24-hour, timezone coordination, patterns \\
    Input Forms & Multi-step forms, validation & 20 & Real-time validation, conditional fields, drag-drop \\
    Nav: Lists \& Tables & Pagination, sorting, filtering & 20 & Infinite scroll, sticky headers, grouping \\
    Nav: Hierarchical & Trees, menus, breadcrumbs & 20 & Expansion states, nested navigation \\
    Nav: Search & Spatial controls, shortcuts & 16 & Map interaction, grid navigation, modals \\
    \midrule
    \textbf{Total} & & \textbf{136} & \\
    \bottomrule
  \end{tabular}
\end{table*}

Simple scenarios systematically cover fundamental UI interactions. Control distribution emphasizes interactive elements: buttons (22.4\%), selection controls (16.6\%), text inputs (13.3\%), and menus (6.6\%). Calendar (8.2\%) and date input controls (6.9\%) together receive notable coverage due to their temporal complexity across 20 distinct implementations including analog clocks and timezone-aware selections.

\begin{figure}[h]
  \centering
  \includegraphics[width=1.0\linewidth]{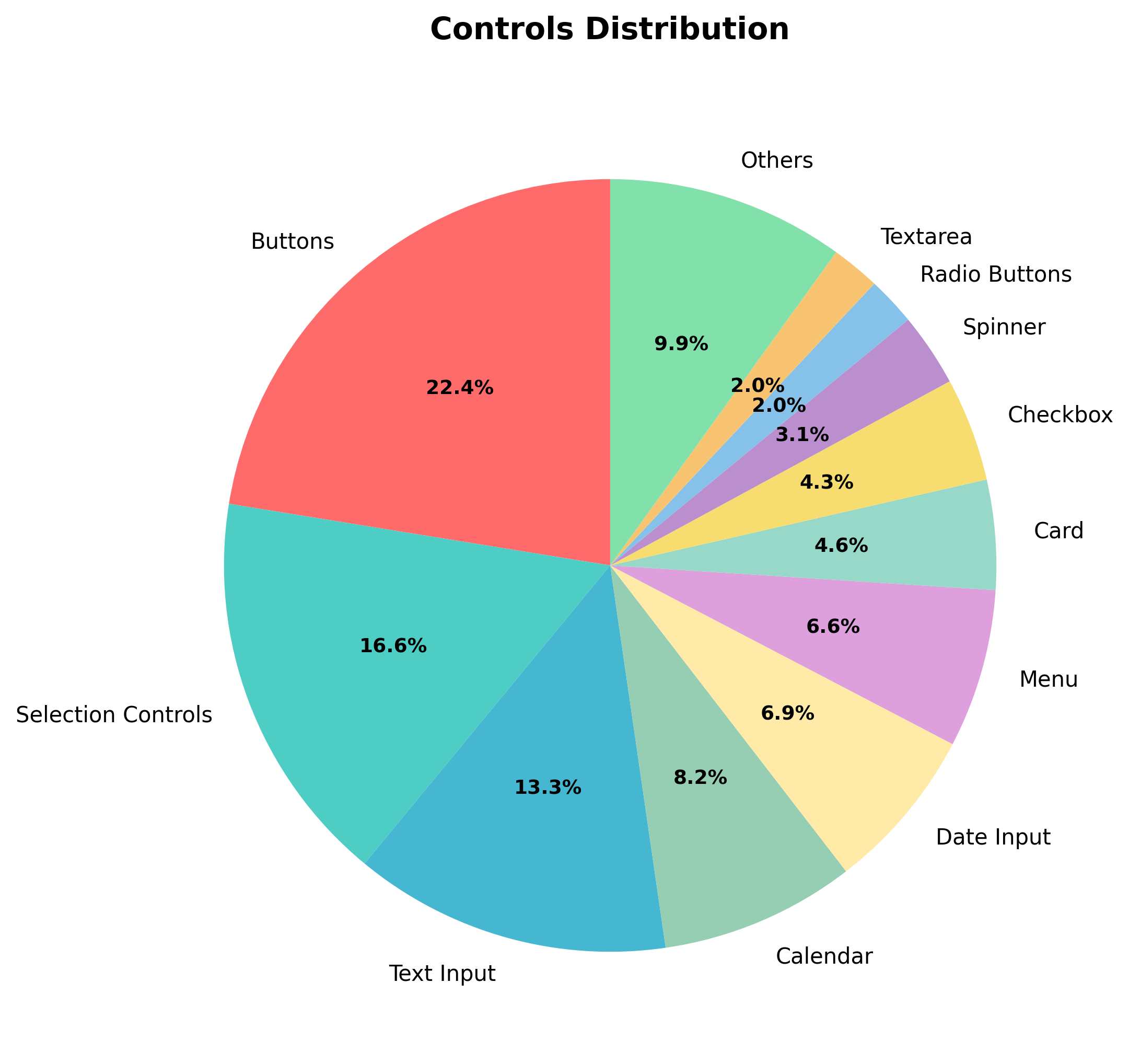}
  \caption{Control distribution across simple scenarios showing systematic coverage of interactive elements.}
  \Description{Pie chart showing distribution of UI controls: Buttons (22.4\%), Selection Controls (16.6\%), Text Input (13.3\%), Calendar (8.2\%), Date Input (6.9\%), Menu (6.6\%), Card (4.6\%), Checkbox (4.3\%), Spinner (3.1\%), Radio Buttons (2.0\%), Textarea (2.0\%), and others (9.9\%).}
  \label{fig:controls-distribution}
\end{figure}

Rather than sampling interface patterns randomly, we target specific interaction challenges. Date/time controls prove particularly demanding, with implementations spanning digital inputs, analog displays, and timezone conversion logic. Combo box interactions receive focused attention through 20 archetypes testing autocomplete-first designs, browse-heavy dropdowns, and token-based multi-selection patterns.

For each control type, we first identified systematic variation dimensions, then implemented 15-20 variants covering that design space. For example, our 20 date picker tasks systematically explore eight dimensions of temporal selection interfaces: selection mode (single, range, multi, recurring), interaction pattern (click, typing, dropdown, timeline, touch, keyboard), UI layout (popup, inline, modal, compact), quick access (presets, contextual suggestions, AI-assisted), business rules (date restrictions, validation), accessibility (keyboard-only, screen reader, mobile), time/timezone integration, and performance/scale (large date ranges, lazy loading). Combo boxes similarly span eight dimensions, varying by trigger mechanism---whether focus, click, or typing activates the dropdown---by input constraints from strict list-only to permissive freeform, and by filtering algorithm from simple substring matching to fuzzy non-contiguous character matching. This approach yielded systematic coverage across control types.

\begin{figure}[h]
  \centering
  \includegraphics[width=1.0\linewidth]{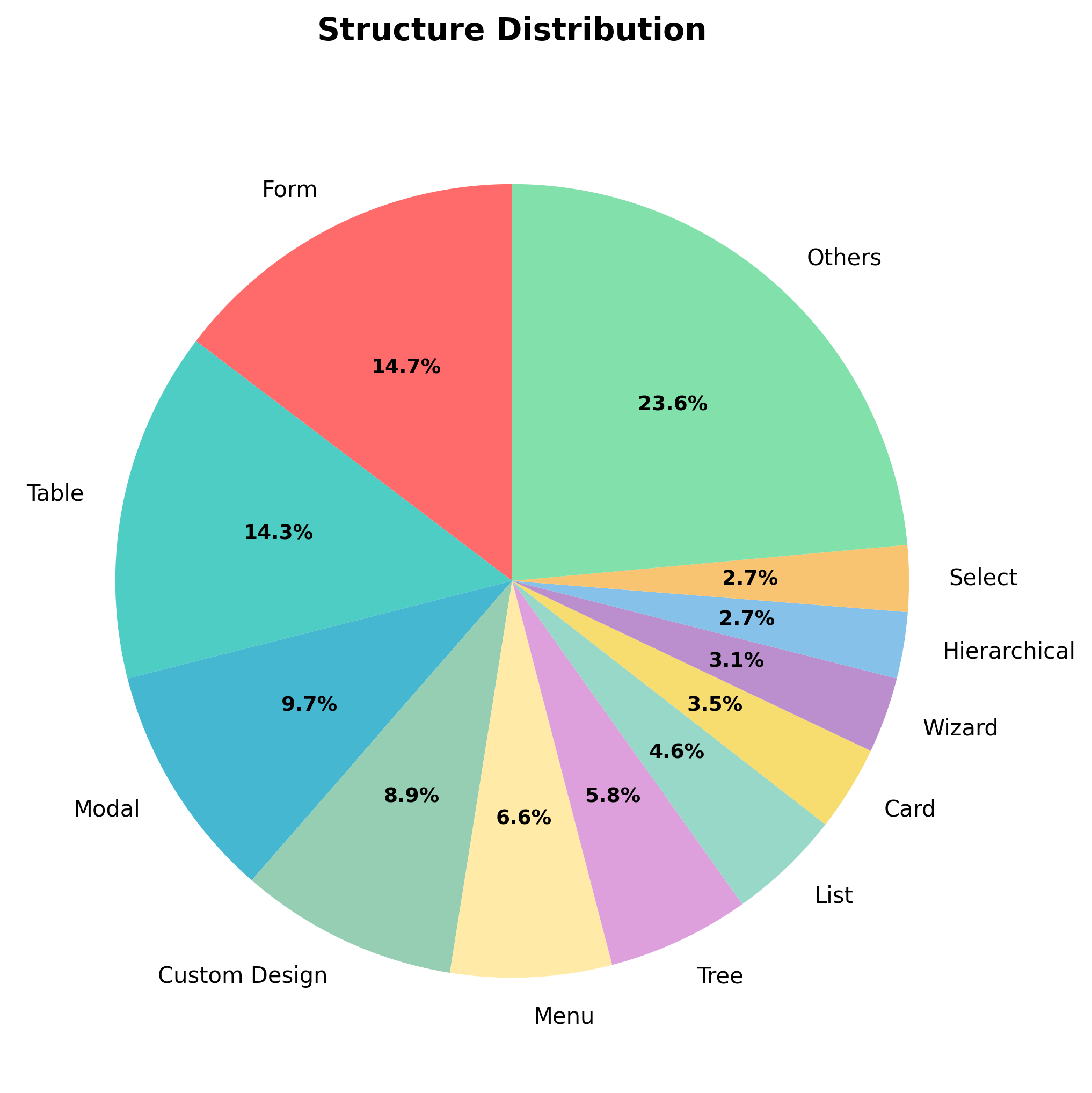}
  \caption{Structure distribution showing layout complexity with emphasis on forms, tables, and custom designs.}
  \Description{Pie chart showing UI structure distribution: Form (14.7\%), Table (14.3\%), Modal (9.7\%), Custom Design (8.9\%), Menu (6.6\%), Tree (5.8\%), List (4.6\%), Card (3.5\%), Wizard (3.1\%), Hierarchical (2.7\%), Select (2.7\%), and others (23.6\%).}
  \label{fig:structure-distribution}
\end{figure}

\begin{figure}[h]
  \centering
  \includegraphics[width=1.0\linewidth]{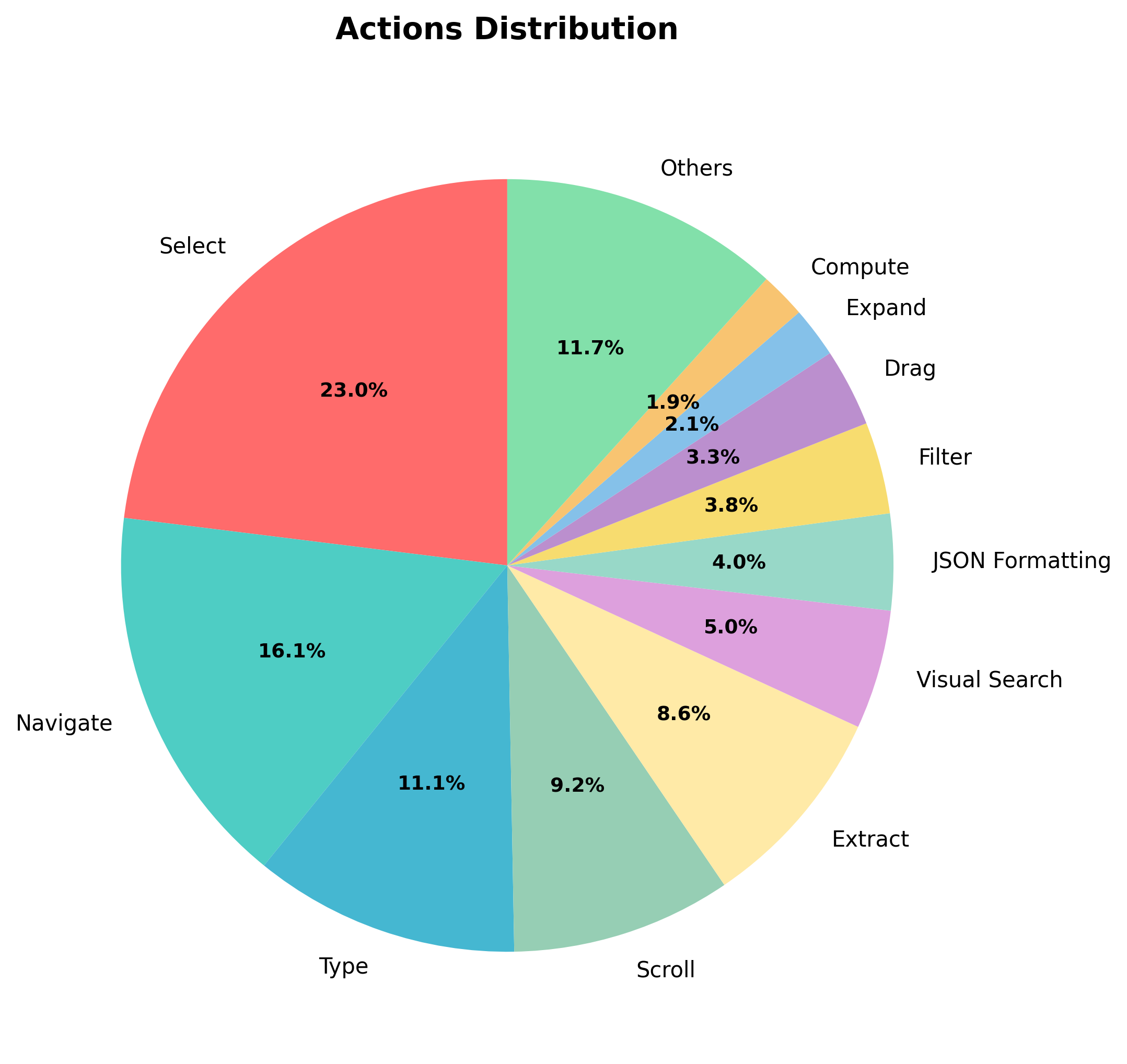}
  \caption{Action distribution reflecting real-world automation patterns with selection operations dominating enterprise workflows.}
  \Description{Pie chart showing action distribution: Select (23\%), Navigate (16.1\%), Type (11.1\%), Scroll (9.2\%), Extract (8.6\%), Visual Search (5\%), JSON Formatting (4\%), Filter (3.8\%), Drag (3.3\%), Expand (2.1\%), Compute (1.9\%), and others (11.7\%).}
  \label{fig:actions-distribution}
\end{figure}

After implementation, we empirically classified the resulting 226 tasks along three dimensions to measure achieved coverage: 22 distinct control types, 27 structure types, and 27 action types. The observed distributions, shown in Figures~\ref{fig:controls-distribution}, \ref{fig:structure-distribution}, and \ref{fig:actions-distribution}, confirm comprehensive coverage of interface patterns agents encounter in real applications.

Structure distribution reveals the layout complexity agents must navigate. Forms (14.7\%) and tables (14.3\%) dominate, reflecting the prevalence of data entry and tabular manipulation in enterprise workflows. Modals (9.7\%) and custom designs (8.9\%) follow, testing adaptation to layered or unconventional layouts. Menus (6.6\%), trees (5.8\%), and lists (4.6\%) capture hierarchical and sequential navigation, while cards (3.5\%) and wizards (3.1\%) test structured multi-step interactions. Select controls (2.7\%) and hierarchical structures (2.7\%) round out the coverage of common enterprise layout patterns.

Action distribution patterns expose the motor complexity of realistic automation. Selection dominates at 23%, followed by navigation and typing. Beyond these mechanical operations, tasks require perceptual skills---extraction and visual search---and cognitive transformations---JSON formatting, filtering, and computation. This mix between mechanical selection and cognitive transformation reflects real enterprise automation scenarios where agents must both manipulate interfaces and process information.

\subsection{Copy-Paste and Iteration Tasks}

\begin{table*}
  \caption{Copy-Paste and Business Process Task Distribution}
  \label{tab:intermediate-tasks}
  \begin{tabular}{p{3.5cm}p{3cm}cp{6.5cm}}
    \toprule
    Task Collection & Activity Type & Task Count & Examples\\
    \midrule
    \multirow{7}{*}{Copy-Paste Tasks} & Data Aggregation \& Collection & 7 & Contact details from tables, quarterly sales compilation, calendar meetings, medical lab results, appointment scheduling \\
    & Filtering \& Search & 6 & Product catalog browsing, urgent task compilation, inventory analysis, library catalog search, electronics filtering \\
    & Content Analysis \& Validation & 6 & Invoice analysis, spam detection, survey counting, resource conflicts, document diff review, accessibility audit \\
    & External Reference \& Form Processing & 9 & Customer registration wizard, service tickets, library cataloging, student enrollment, tenant application, loan application \\
    & Classification \& Rule-Based Processing & 6 & Product configuration validator, code review resolution, service taxonomy mapping, design annotation validation \\
    & Data Manipulation & 3 & Salary updates, Kanban ticket management, social media scheduling \\
    & Navigation \& Exploration & 3 & File tree PDF collection, org chart traversal, menu hierarchy mapping \\
    \midrule
    \multirow{6}{*}{Business Process Tasks} & Error Recovery \& Batch Processing & 2 & Connection timeout handling, file access with error navigation \\
    & Data Quality \& Transformation & 2 & Currency normalization, customer deduplication \\
    & Form Management \& Data Entry & 3 & Accordion form completion, multi-step transport form, paginated table extraction \\
    & Compliance \& PII Management & 1 & Customer record anonymization \\
    & Application Review & 1 & Graduate school application assessment \\
    & Directory Analysis & 1 & Transport company fleet counting \\
    \midrule
    \textbf{Total} & & \textbf{50} & \\
    \bottomrule
  \end{tabular}
\end{table*}

The copy-paste and business process tasks bridge atomic actions and workflow complexity with 50 tasks. Action patterns shift toward coordination: navigation increases relative to simple selection, with substantial scrolling and filtering requirements. Structure distribution emphasizes data-heavy layouts: tables, hierarchical trees, and complex custom designs.

Tasks require synthesizing information across sources: Excel-to-form mapping, quarterly sales compilation, and organizational chart navigation. \textbf{Memory challenges} emerge naturally---agents must track processed items, maintain running totals across pagination, and avoid duplicate selections while navigating complex interfaces.

Our design centers on three cognitive challenges that current CUAs must master for reliable enterprise deployment. \textbf{Aggregation and correlation patterns} require agents to synthesize information across multiple sources while maintaining relationships between disparate data points---collecting contact details from scattered table rows, correlating quarterly sales figures across separate databases, or extracting director-level employees from organizational charts based on hierarchical position.

\textbf{Conditional execution} logic appears throughout tasks requiring rule-based reasoning within interactive contexts: updating salaries for employees hired before specific dates, scheduling social media posts exceeding engagement thresholds, or flagging document changes against compliance rules. \textbf{Memory and state management} challenges test whether agents can maintain working context across extended interactions: tracking which invoices have been processed while navigating pagination, remembering expansion states in hierarchical trees, or avoiding duplicate item selection during systematic interface exploration.

Business logic appears throughout: temporal filtering (employees hired before 2020), performance thresholds (social media posts with $>$100 likes), and compliance validation (document changes against style guides). Beyond explicit conditional logic, tasks include scraping and triage scenarios requiring visual or semantic inferences about content properties. Implementation scales appropriately: simple tasks require $\sim$100 lines of code, copy-paste scenarios demand $\sim$400 lines to model realistic data relationships.

\subsection{Enterprise Applications}

\begin{table*}[!htbp]
  \caption{Enterprise Application Task Distribution}
  \label{tab:enterprise-apps}
  \begin{tabular}{p{3cm}cp{4cm}p{6cm}}
    \toprule
    Application & Task Count & Primary Workflows & Complexity Areas\\
    \midrule
    Salesforce (CRM) & 8 & Lead management, activity tracking, bulk operations & Multi-step lead creation, cross-record activity transfers, search/sort/filter operations, bulk status updates \\
    SAP Stock Overview (ERP) & 8 & Inventory management, material lookup, stock analysis & Complex filtering with multiple criteria, hierarchical data navigation, cross-plant material tracking \\
    Concur (Expense Management) & 8 & Expense reporting, receipt attachment, vendor management, flight booking & Multi-step report creation, document attachment workflows, conditional business logic, travel search interfaces \\
    Workday (HR Management) & 8 & Time-off requests, contact management, personal information updates & Date range selection, hierarchical menu navigation, form validation workflows \\
    Kanban Board (Project Management) & 8 & Issue tracking, status management, project coordination & Task lifecycle management, drag-drop operations, view switching, data extraction \\
    \midrule
    \textbf{Total} & \textbf{40} & & \\
    \bottomrule
  \end{tabular}
\end{table*}

Enterprise scenarios use faithful mocks of key workflows from Salesforce, SAP, Workday, Concur, and Kanban systems, focusing on representative UI patterns and business logic rather than full application coverage. Structure distribution reflects enterprise complexity: forms (14.7\%) and tables (14.3\%) define the core layout, while custom designs (8.9\%), modals (9.7\%), and error states (2.7\%) capture the irregular conditions common in enterprise software.

Application mocking preserves essential complexity through authentic business workflows and enterprise UI patterns while maintaining reproducibility. Each mock incorporates realistic domain logic: multi-step approval processes, conditional form validation, and hierarchical data navigation. Implementation cost scales primarily with application architecture rather than individual task complexity---substantial codebases (2000-4000 lines per application) reflect enterprise software's inherent UI and state management requirements, while individual tasks typically require 100-300 lines including validation and guidance systems. This architecture enables testing both simple operations (lead creation, data lookup) and complex multi-step workflows (expense report creation with attachments) within the same realistic enterprise context. The benchmark's strength lies in providing authentic enterprise environments for agent evaluation rather than artificially inflating task difficulty.

\subsection{Multi-Resolution Testing}

All tests execute across three enforced resolutions: 1024$\times$768 (XGA), Full HD (1080p), and 4K, reflecting real-world deployment variability. This addresses environmental brittleness, a failure mode typically overlooked in existing benchmarks. Our empirical results show that screen resolution has a non-trivial impact on agent performance---as resolution increases, agents exhibit higher frequency of grounding errors, often predicting incorrect element positions due to the larger action space. While higher resolution can reduce exploratory actions by making more interface elements visible simultaneously, the overall effect is performance degradation, suggesting current agents struggle to generalize their perception and grounding mechanisms across different screen scales.

\section{Environment Architecture}

To enable reproducible agent evaluation, UI-CUBE provides a containerized environment where agents interact with realistic enterprise UIs through standardized interfaces. This section describes the evaluation infrastructure, how agents integrate with the system, what they observe and control, and why we made specific architectural choices.

\subsection{Evaluation Infrastructure}

Our benchmark runs inside Docker containers, each providing an isolated Ubuntu desktop environment with a browser displaying the task interface. Containers expose three access points: a VNC server for visual observation and input control, Chrome DevTools Protocol for DOM inspection and validation, and an HTTP server delivering the benchmark application. This design supports both pure vision-based agents that work from screenshots alone, and multimodal agents that combine visual and semantic understanding.

The system handles parallel evaluation automatically. Multiple containers can run simultaneously on a single machine, with the orchestration layer managing port allocation to avoid conflicts. A typical 16GB server supports around 15 concurrent evaluation instances, enabling efficient large-scale testing. Containers start quickly---under 20 seconds from launch to ready state---making iterative development practical.

We chose containers over virtual machines for two reasons: density and speed. Containers share the host kernel, allowing more instances per machine and faster initialization compared to full VMs. Since our benchmark focuses on application-level interactions rather than OS-level operations, container isolation provides sufficient separation while dramatically improving evaluation throughput.

\subsection{Agent Integration Protocol}

Agents integrate with UI-CUBE by implementing the \texttt{Agent} base class, which provides standardized methods for environment interaction, observation gathering, and execution control. The core integration requires implementing a single abstract method:

\begin{lstlisting}[language=Python, caption=Agent integration interface, basicstyle=\ttfamily\footnotesize]
from uitask.computer_use import Agent

class CustomAgent(Agent):
    def __init__(self, task, output_dir, vnc_port, cdp_port):
        super().__init__(output_dir, vnc_port=vnc_port, 
                         cdp_port=cdp_port)
        self.task = task
        self.model = initialize_model()
        
    def act(self) -> bool:
        # Get observation
        screenshot = self.screenshot()  # base64 PNG
        
        # Model inference
        action = self.model.predict(screenshot, self.task)
        
        # Execute action
        if action.type == "click":
            self.mouse_click(MouseClickAction(
                button=MouseButton.LEFT,
                position=Position(action.x, action.y)
            ))
        elif action.type == "type":
            self.type_text(action.text)
            
        # Check completion
        return self.check_task_success()
\end{lstlisting}

The \texttt{act()} method executes a single agent step, returning \texttt{True} when the task is complete. The base class provides the \texttt{run(max\_steps)} harness that iteratively calls \texttt{act()} until completion or step exhaustion, automatically managing recording lifecycle, action history tracking, and error handling. This design separates agent reasoning logic from environment management, enabling researchers to focus on model development rather than infrastructure.

The system automatically records all agent actions to enable fault tolerance and post-execution analysis. If an agent crashes mid-run, the partial trajectory file preserves all actions up to the failure point. Complete runs produce full execution traces annotated with task completion status at each step, supporting detailed performance analysis and debugging.

\subsection{Observation Space}

Agents observe the environment through three complementary channels, supporting different architectural approaches.

Vision-based agents work primarily from screenshots---full desktop captures returned as base64-encoded images. These show everything visible on screen, including the cursor position, exactly as a human would see it. The VNC protocol transmits only changed screen regions between frames, making continuous observation efficient.

Agents can also inspect the DOM structure through Chrome DevTools Protocol, accessing semantic information beyond pixels. This enables extracting text content, reading element attributes, and understanding document structure---capabilities useful for agents that combine visual and linguistic reasoning. The same CDP connection allows agents to check task completion status by reading validation sentinels that appear in the DOM when test functions execute.

\subsection{Action Space}

Agents control the environment through eight primitive operations that mirror human computer interaction. Table~\ref{tab:action-space} shows the complete action interface.

\begin{table}[h]
  \caption{Action Space Primitives}
  \label{tab:action-space}
  \begin{tabular}{@{}%
    >{\ttfamily\raggedright\arraybackslash}p{2.6cm}%
    >{\ttfamily\raggedright\arraybackslash}p{\dimexpr\linewidth-2.6cm-2\tabcolsep\relax}@{}}
    \toprule
    Action Method & Parameters\\
    \midrule
    mouse\_move & position: Position \\
    mouse\_click & action: MouseClickAction \\
                & button: MouseButton \\
                & position: Position \textbar{} None \\
                & click\_type: MouseClickType \\
    mouse\_drag  & action: MouseDragAction \\
                & button: MouseButton \\
                & start: Position \\
                & end: Position \\
    mouse\_scroll& action: ScrollAction \\
                & position: Position \\
                & direction: ScrollDirection \\
                & repeat: int \\
    press\_keys  & keys: Sequence[str] \\
    type\_text   & text: str \\
    page\_navigation & direction: Literal["forward","back"] \\
    wait        & duration: float \\
    finish      & params: dict \textbar{} None \\
    \bottomrule
  \end{tabular}
\end{table}

Mouse operations work through pixel coordinates, requiring agents to visually ground elements before interaction. Agents specify exact screen positions for clicks, drags, and scrolls---there is no selector-based clicking or privileged element access. This design tests coordinate-based grounding capabilities essential for real-world deployment where agents lack application-specific APIs.

Keyboard operations handle both individual key presses and text input. The \texttt{press\_keys} method supports modifier combinations like \texttt{Ctrl+Shift+T}, parsing them automatically to hold the appropriate keys simultaneously. The \texttt{type\_text} method handles arbitrary Unicode text, including international characters and emoji. Browser navigation commands provide forward/back controls.

Finally, control operations let agents wait for UI updates to complete and signal task completion. These primitives combine to enable the full range of interactions humans perform---from simple clicks to complex multi-step workflows involving dragging, form filling, and navigation.

Behind the scenes, these high-level actions translate to VNC protocol messages that the system injects into the desktop environment, making them indistinguishable from physical user input. This design allows testing across different application types without requiring application-specific automation tools, though it trades some execution speed for this generality and realism.

\subsection{Design Rationale}

Key architectural choices distinguish UI-CUBE from existing evaluation frameworks, each motivated by enterprise deployment requirements.

The VNC protocol might seem antiquated compared to modern browser automation tools like Playwright or Selenium. However, VNC offers a crucial advantage: universality. The same agent code that controls a web browser can control LibreOffice, desktop applications, or any future additions to the benchmark. VNC also tests coordinate-based grounding---agents must locate elements visually rather than accessing them through privileged APIs. This matches real deployment scenarios where agents lack application-specific integration. The built-in screen recording capabilities further simplify trajectory analysis.

We test every task at three different screen resolutions. This exposes an often-overlooked problem: agents that work at one resolution may fail at another, even though the task is identical. Our results show 10-20\% performance variation across resolutions, revealing brittleness in current agent architectures. Since real users work across diverse hardware---laptops, desktops, tablets---this variability testing matters for deployment readiness.

\section{Execution-Based Evaluation}

Our evaluation methodology centers on programmatic oracles over application state rather than trajectory-based judges. This approach addresses the validation fragility issues identified in existing benchmarks by implementing deterministic, replayable assessment mechanisms that inspect final system state rather than execution paths.

\subsection{State-Diff Validation Architecture}

Each task implements a \texttt{test()} function that inspects application state or user-submitted JSON, returning \texttt{\{success, message\}} based on programmatic assertions. The application state is exposed through the global \texttt{window.app\_state} object.

Examples of \texttt{app\_state} structures:

\begin{lstlisting}[style=jsblock, caption=Kanban app state]
window.app_state = {
  tasks: [
    {
      id: 'ML-2445',
      title: 'Fix model fallback',
      status: 'done',
      assignee: { name: 'Anna Petrov' },
      priority: 'high'
    }
  ],
  operationHistory: [
    { type: 'drag_drop', taskId: 'ML-2445', toStatus: 'done' }
  ],
  currentView: 'kanban',
  filters: { status: 'todo' }
};
\end{lstlisting}

\begin{lstlisting}[style=jsblock, caption=Workday app state]
window.app_state = {
  submittedRequests: [
    {
      id: 'req1',
      absenceType: 'PTO',
      dates: [ { day: 4, month: 7, year: 2025 } ]
    }
  ],
  contactInformation: {
    phones: [ { number: '(555) 123-7654', usage: 'Home' } ]
  }
};
\end{lstlisting}

For enterprise workflow validation, our oracles assert application truth directly through React state inspection. Workday absence requests verify temporal logic by scanning submittedRequests collections, filtering by date ranges, and comparing day arrays against target specifications. Kanban task creation validates entity-level fields---titles, assignees, priorities, project associations---through direct state inspection rather than interface parsing. This approach eliminates brittleness from harmless interface variations while maintaining strict correctness requirements.

\subsection{Multi-Tier Validation Strategy}

Our two-tier benchmark architecture requires correspondingly sophisticated validation approaches. Simple scenarios employ exact equality checks over discrete interface states---button selections, form field values, navigation positions. The deterministic nature of atomic interactions permits strict matching without tolerance bounds.

Complex workflow tasks introduce result-submission validation modes where agents submit structured JSON against expected objects. These tasks require exact matches on strings and sets (via sorted arrays), numeric count validation, and shape correctness guards that reject malformed submissions before content verification. The validation preserves determinism while accommodating the data transformation nature of workflow tasks.

Enterprise application validation broadens oracle expressivity to handle complex business processes. Long-horizon behaviors require state persistence validation---resumption after timeouts, batch processing completion, error recovery sequences. Numeric normalization tasks employ regex-based leniency for format variations while maintaining semantic correctness. Navigation tasks validate goal achievement with minimum interaction thresholds rather than exact step counts.

\subsection{Validation Robustness and Brittleness}

Our oracle design balances exactness with operational robustness. Strict equality checks ensure unambiguous correctness assessment, critical for benchmark reliability. Format sensitivity brittleness is mitigated through systematic normalization and order-independence where semantically appropriate. Whitespace trimming, case-insensitive comparisons, and array sorting ensure identical semantic results are recognized regardless of presentation variations that don't affect business logic.

Semi-fuzzy validation emerges in analytical tasks requiring structured reasoning. Content categorization employs bounded error tolerance ($\leq$1 error per classification set) while maintaining semantic accuracy requirements. Extraction tasks validate against precomputed ground-truth mappings with controlled flexibility for equivalent expressions.

This validation architecture establishes a domain-specific language by convention where oracles function as pure functions over application state---deterministic, idempotent, and auditable. The approach enables systematic capability assessment across interface complexity tiers while maintaining the evaluation rigor required for enterprise deployment readiness assessment.

\subsection{Implementation Characteristics}

Task implementation scales predictably across complexity tiers. Simple scenarios require approximately 100 lines including validation logic, complex workflow tasks demand 400 lines to model realistic data relationships and state transitions, while enterprise applications require substantial validation frameworks (100-300 lines per task) to handle complex business logic and error conditions.

The oracle architecture separates validation logic from interface implementation, enabling identical correctness criteria across resolution variations and interface mutations. This separation proves essential for systematic evaluation, as identical business requirements can be validated consistently despite environmental variations that typically introduce evaluation noise in existing benchmarks.

\section{Results and Evaluation}

\subsection{Model Performance}

We evaluated five state-of-the-art computer use agents on UI-CUBE. Claude Computer Use 4.0 and OpenAI-computer-use-preview are end-to-end AI agents designed to interact with computer systems through natural language commands. Our UIPathScreenAgent adapts the observe-reason-act paradigm of ReAct~\cite{yao2022react} through a hierarchical architecture: a large language model planner reasoning about task goals and generating action sequences, and a visual grounder combining UI-TARS-7b~\cite{qin2025uitars} with the UiPath computer vision model to translate actions into precise screen coordinates. Task and action reviewer components maintain the feedback loop by analyzing execution results and triggering replanning when needed. We evaluate UIPathScreenAgent with three foundation models as planners: Gemini 2.5 Flash~\cite{comanici2025gemini}, GPT-5 mini, and GPT-5.

\begin{table*}[h]
  \caption{Average Performance Across All Resolutions}
  \label{tab:average-performance}
  \begin{tabular}{lcc}
    \toprule
    Model & Simple Tasks & Complex Tasks\\
    \midrule
    Claude Computer Use 4.0 & 66.7\% & 9.5\% \\
    OpenAI-computer-use-preview & 70.3\% & 10.5\% \\
    UIPathScreenAgent/Gemini 2.5 Flash & 68.6\% & 11.9\% \\
    UIPathScreenAgent/GPT-5 mini & 77.0\% & 18.4\% \\
    UIPathScreenAgent/GPT-5 & 84.8\% & 19.4\% \\
    \bottomrule
  \end{tabular}
\end{table*}

\begin{table*}[h]
  \caption{Performance by Resolution}
  \label{tab:resolution-performance}
  \begin{tabular}{lcccccc}
    \toprule
    Model & \makecell{Simple\\XGA} & \makecell{Complex\\XGA} & \makecell{Simple\\1080p} & \makecell{Complex\\1080p} & \makecell{Simple\\4K} & \makecell{Complex\\4k}\\
    \midrule
    Claude Computer Use 4.0 & 89.0\% & 11.9\% & 73.5\% & 11.9\% & 37.5\% & 4.5\% \\
    OpenAI-computer-use-preview & 91.2\% & 11.9\% & 84.6\% & 16.4\% & 35.3\% & 3.0\% \\
    UIPathScreenAgent/Gemini 2.5 Flash & 84.6\% & 10.5\% & 82.4\% & 13.4\% & 39.0\% & 11.9\% \\
    UIPathScreenAgent/GPT-5 mini & 90.4\% & 13.4\% & 89.0\% & 26.9\% & 51.5\% & 14.9\% \\
    UIPathScreenAgent/GPT-5 & 94.1\% & 17.9\% & 91.2\% & 28.4\% & 69.1\% & 11.9\% \\
    \bottomrule
  \end{tabular}
\end{table*}

The results reveal a sharp capability cliff rather than gradual performance degradation. While agents achieve reasonable performance on simple UI interactions (67-85\% averaged across resolutions), complex tasks show precipitous drops to 9-19\% success rates. This discontinuous performance pattern indicates fundamental architectural limitations rather than incremental capability differences.

Our findings align with recent observations in WorkArena++ \cite{boisvert2024workarena}, which demonstrated a similar capability cliff: GPT-4o achieved 42.7\% on atomic tasks (L1) but dropped precipitously to 3\% on compositional workflows (L2) and 0\% on complex reasoning tasks (L3), while humans maintained 93.9\% success after minimal training. CRMArena-Pro~\cite{huang2025crmarenapro} independently validates this pattern in CRM workflows: leading agents achieve 58\% in single-turn scenarios but degrade to 35\% in multi-turn settings.

This consistent pattern across UI-CUBE (67-85\% simple $\rightarrow$ 9-19\% complex), WorkArena++ (42.7\% $\rightarrow$ 3\%), and CRMArena-Pro (58\% $\rightarrow$ 35\%) suggests the performance degradation stems from limitations in sustained memory and state coordination rather than difficulties with individual reasoning steps.

The key insight emerges from examining human-agent performance ratios: on simple tasks, agents achieve 68-87\% of human performance, but this ratio collapses to 15-32\% on complex workflows. If the gap were merely a scaling challenge, we would expect agents to maintain proportional performance relative to humans across complexity tiers. Instead, the disproportionate degradation reveals fundamental limitations in memory management, hierarchical planning, and state coordination that distinguish current agent architectures from human cognitive capabilities.

Performance degrades significantly as resolution increases for most models, particularly at 4K resolution where simple task performance drops by 40-55 percentage points for Claude and OpenAI models compared to XGA performance. This suggests current agents struggle to generalize their perception and grounding mechanisms across different screen scales, despite the potential benefit of having more interface elements visible simultaneously.

\subsection{Human Baseline}

To establish realistic performance ceilings, we conducted comprehensive human evaluations across all task collections. Human evaluators with no prior experience with the benchmark applications were given task instructions and asked to complete them manually through the web interface.

\begin{table*}[h]
  \caption{Human Performance by Task Collection}
  \label{tab:human-performance-detailed}
  \begin{tabular}{lccc}
    \toprule
    Collection & Accuracy & Avg. Duration (s) & Avg. Steps\\
    \midrule
    \multicolumn{4}{l}{\textit{Simple Tasks}}\\
    Combo box tasks & 100.0\% & 11.0 & 4.1\\
    Date pickers & 100.0\% & 15.3 & 7.2\\
    Time pickers & 100.0\% & 9.5 & 6.0\\
    Input boxes & 90.0\% & 20.6 & 11.3\\
    Navigation: lists/tables & 95.0\% & 6.8 & 4.3\\
    Navigation: hierarchical & 100.0\% & 2.9 & 2.1\\
    Navigation: search & 100.0\% & 8.7 & 4.3\\
    Kanban board & 100.0\% & 12.5 & 3.8\\
    \midrule
    \multicolumn{4}{l}{\textit{Complex Tasks}}\\
    Business process tasks & 50.0\% & 267.2 & 75.4\\
    Copy-paste tasks & 53.3\% & 177.7 & 72.9\\
    Salesforce (CRM) & 62.5\% & 122.5 & 55.5\\
    Concur (Expense) & 50.0\% & 130.4 & 34.1\\
    SAP Stock (ERP) & 50.0\% & 183.3 & 33.6\\
    Workday (HR) & 62.5\% & 42.8 & 14.3\\
    \bottomrule
  \end{tabular}
\end{table*}

Table~\ref{tab:human-performance-detailed} reveals striking performance differences between task categories. Simple tasks achieve near-perfect performance (90-100\%), with most interface interactions completed in under 20 seconds. However, complex tasks show dramatic performance degradation (50-62.5\% accuracy), with execution times increasing 10-20x.

\begin{table}[h]
  \caption{Human Performance Summary}
  \label{tab:human-performance-summary}
  \begin{tabular}{lccc}
    \toprule
    Task Type & Tasks & Accuracy & Avg. Duration (s)\\
    \midrule
    Simple Tasks & 136 & 97.9\% & 10.7\\
    Complex Tasks & 90 & 61.2\% & 149.2\\
    \bottomrule
  \end{tabular}
\end{table}

Overall, simple tasks achieve near-perfect performance (97.9\%) with rapid completion, while complex tasks show only 61.2\% accuracy despite significantly longer execution times.

Several factors contribute to lower performance on complex tasks: (1) Task description complexity involving multi-step workflows with intricate business logic requiring careful reading and comprehension; (2) Manual JSON formatting prone to syntax errors and requiring precise attention to data structure requirements; (3) Long execution sequences (150+ steps) with strict validation where a single mistake invalidates the entire execution; (4) Lack of prior application experience with enterprise systems like Salesforce, SAP, or Workday, requiring learning unfamiliar interfaces on-the-fly.

These human baseline results establish that the 61.2\% accuracy on complex tasks represents a realistic ceiling for current AI systems operating in enterprise environments, particularly when dealing with unfamiliar applications and multi-step workflows requiring sustained attention and precise execution.

\subsection{Agent Efficiency Analysis}

Beyond success rates, we analyze agent efficiency by comparing steps required for task completion against human baselines. This metric reveals execution overhead independent of accuracy, exposing inefficiencies in navigation, decision-making, and action selection.

\begin{table*}[h]
  \caption{Human vs Agent Step Efficiency at 1080p}
  \label{tab:step-efficiency}
  \begin{tabular}{lccc}
    \toprule
    Model & \makecell{Simple Tasks\\(Agent/Human)} & \makecell{Complex Tasks\\(Agent/Human)} & \makecell{Overall\\Ratio}\\
    \midrule
    UIPathScreenAgent/Gemini 2.5 Flash & 1.5$\times$ & 1.2$\times$ & 1.4$\times$\\
    OpenAI-computer-use-preview & 1.9$\times$ & 1.2$\times$ & 1.6$\times$\\
    UIPathScreenAgent/GPT-5 & 1.6$\times$ & 1.7$\times$ & 1.7$\times$\\
    UIPathScreenAgent/GPT-5 mini & 1.8$\times$ & 1.4$\times$ & 1.6$\times$\\
    Claude Computer Use 4.0 & 3.3$\times$ & 2.1$\times$ & 2.7$\times$\\
    \bottomrule
  \end{tabular}
\end{table*}

Table~\ref{tab:step-efficiency} shows agents require 1.5--3.3$\times$ more steps than humans on simple tasks (human baseline: 5.6 steps) and 1.2--2.1$\times$ more steps on complex tasks (human baseline: 41.4 steps). These ratios align with OSWorld-Human's findings that leading agents take 1.4--2.7$\times$ more steps than necessary~\cite{abhyankar2025osworld}, validating that step inefficiency is a systemic challenge across computer-use benchmarks rather than an artifact of specific task designs. MMBench-GUI~\cite{wang2025mmbench} further emphasizes efficiency as a critical but underexplored dimension, identifying that all models suffer from substantial inefficiencies with excessive redundant steps, reinforcing that efficiency improvements require architectural rather than parametric advances.

The efficiency patterns reveal architectural insights beyond raw success rates. On simple tasks, Gemini 2.5 Flash achieves the lowest overhead (1.5$\times$), while Claude shows the highest (3.3$\times$), suggesting fundamental differences in visual grounding and action planning. Interestingly, the most efficient models on simple tasks (Gemini, OpenAI) also perform efficiently on complex tasks (1.2$\times$), indicating that core action efficiency transfers across task complexity.

Claude's particularly high step overhead (2.7$\times$ overall) correlates with its lower success rates, suggesting execution inefficiency compounds into task failure. In contrast, Gemini's tight step efficiency (1.4$\times$ overall) demonstrates that architectural designs can minimize unnecessary actions without sacrificing generality---an important consideration for real-world deployment where compute costs and user wait times scale with action counts.

\section{Error Analysis}

Our analysis of agent performance across enterprise applications showed several sources of failure:

\textbf{Lack of internal knowledge about application:} Agents frequently fail to identify the correct navigation icon or correct menu item, particularly when this requires prior knowledge of the interface or exploratory actions such as hovering over elements. This is particularly true for unpopular applications, where the meaning of different UI elements (particularly buttons with icons and no text labels) is not known beforehand.

\textbf{Grounding errors:} Grounding models sometimes produce small coordinate misalignments even if the target UI element is simple to locate without confusion, leading to execution failure. Calendar interfaces prove particularly problematic, with models clicking correct day numbers in wrong months or targeting whitespace adjacent to date cells. Partial visibility of interface elements (element occlusion) often prevents accurate selection.

\textbf{Reasoning errors:}
\begin{itemize}
\item \textbf{Long execution context:} Many enterprise dynamic or iterative tasks involve repetitive procedures applied to multiple items, and agents often lose track of progress within the sequence, either repeating prior steps or omitting elements entirely.
\item \textbf{Unexpected UI changes:} Agents often become stuck when encountering unexpected user interface behavior, particularly when the canonical execution path is unavailable. While exploration is sometimes necessary to recover from such situations, agents lack robust strategies for \textit{safe exploration}, i.e., exploring without performing potentially irreversible actions.
\item \textbf{Scrolling down bias:} Sometimes in the middle of execution, the relevant content requires upward navigation while the agents only try scrolling down.
\end{itemize}

\textbf{Hallucination:} Agents introduce content that does not faithfully reflect the source or the intended task. In copy-paste workflows, agents sometimes "correct" or reformat the input, thereby altering its meaning. In form filling tasks, when fields are left empty, agents may invent plausible but spurious values rather than preserving the emptiness.

\begin{figure}[htbp]
    \centering
    \includegraphics[width=1.0\columnwidth]{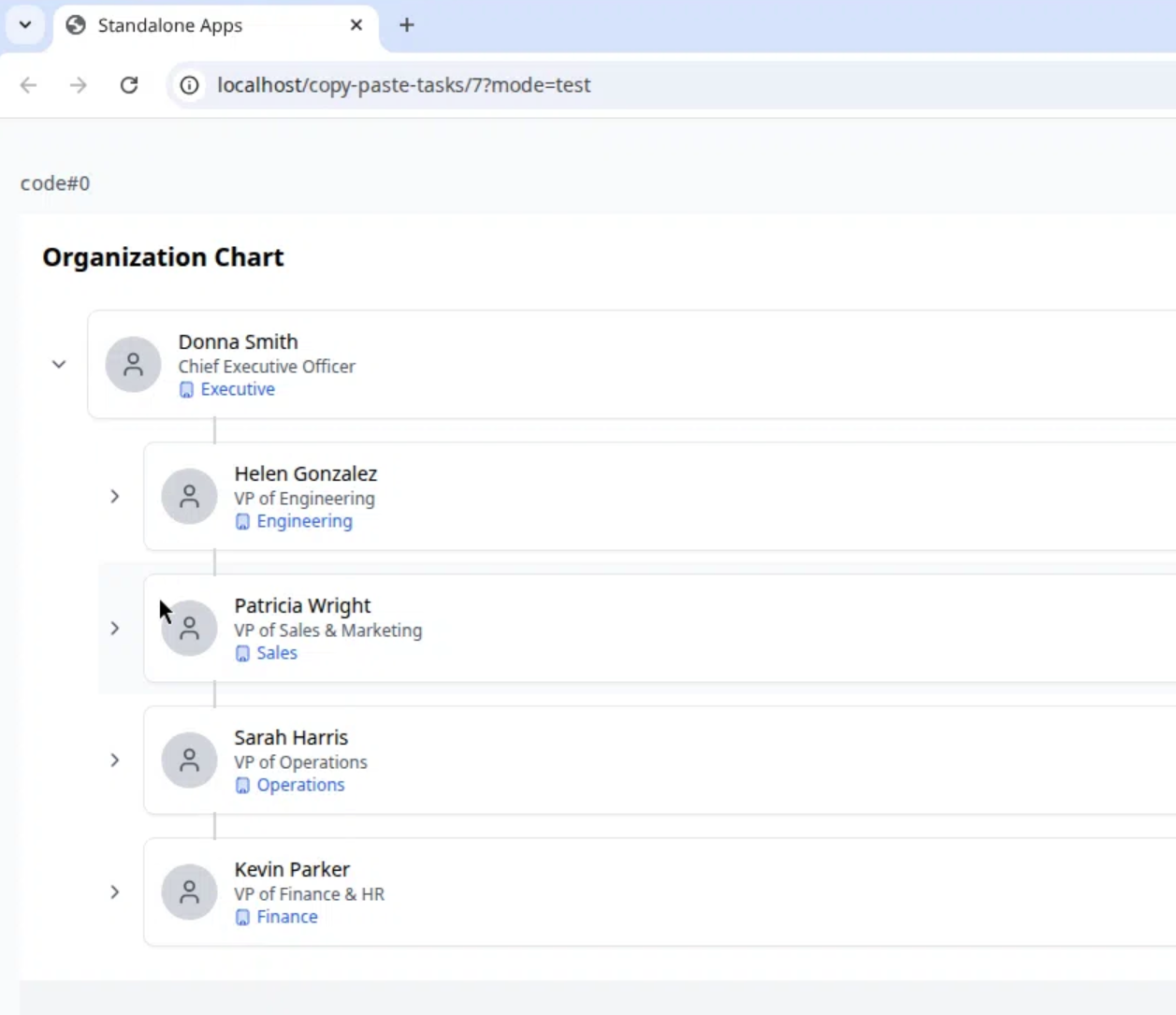}
    \caption{Agent miss-clicks the expander button, demonstrating grounding precision failures in hierarchical navigation contexts.}
    \label{fig:misclick}
\end{figure}

\begin{figure}[htbp]
    \centering
    \includegraphics[width=1.0\columnwidth]{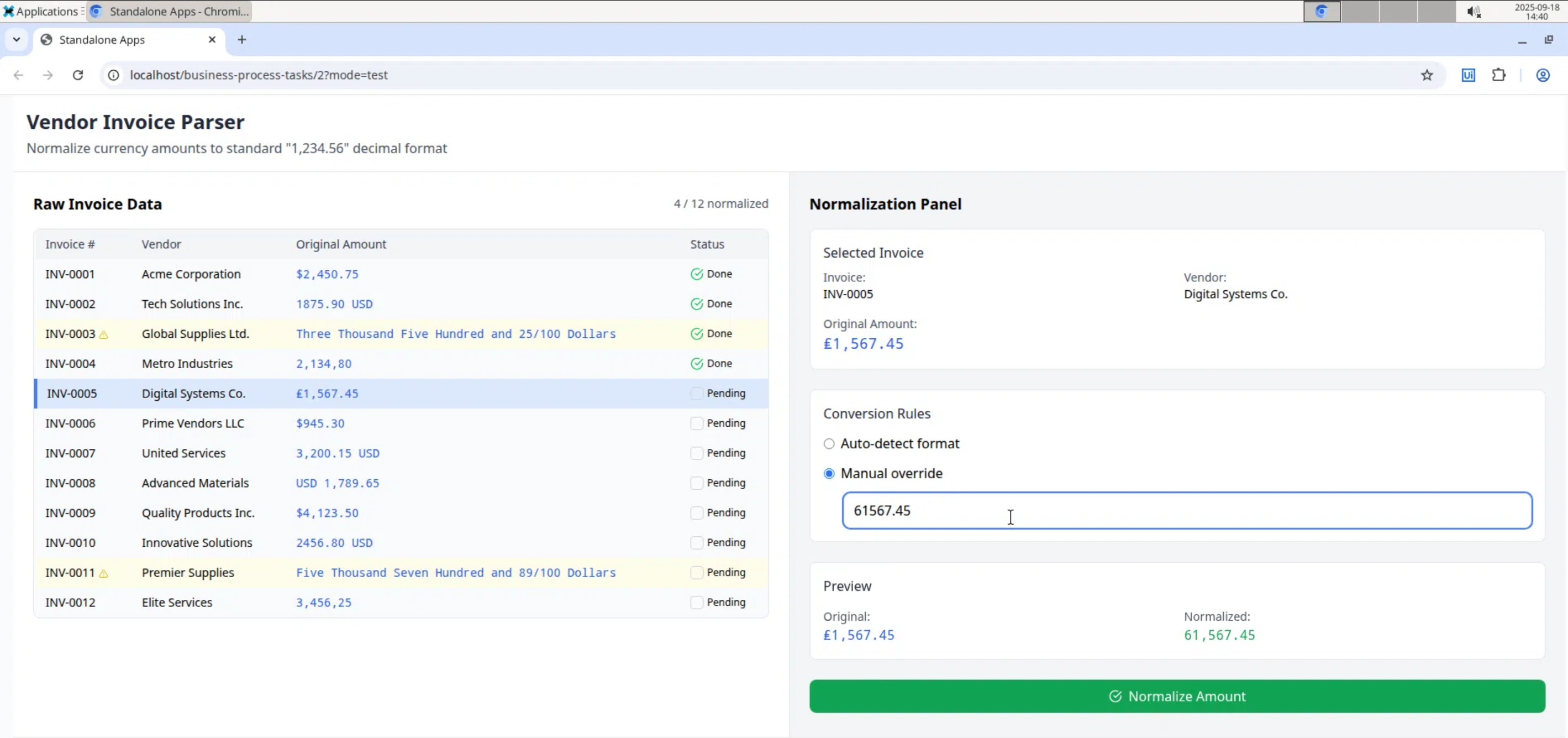}
    \caption{Agent incorrectly reads text from a list, showing perception errors in structured data environments.}
    \label{fig:incorrect-extraction}
\end{figure}

\begin{figure}[htbp]
    \centering
    \includegraphics[width=1.0\columnwidth]{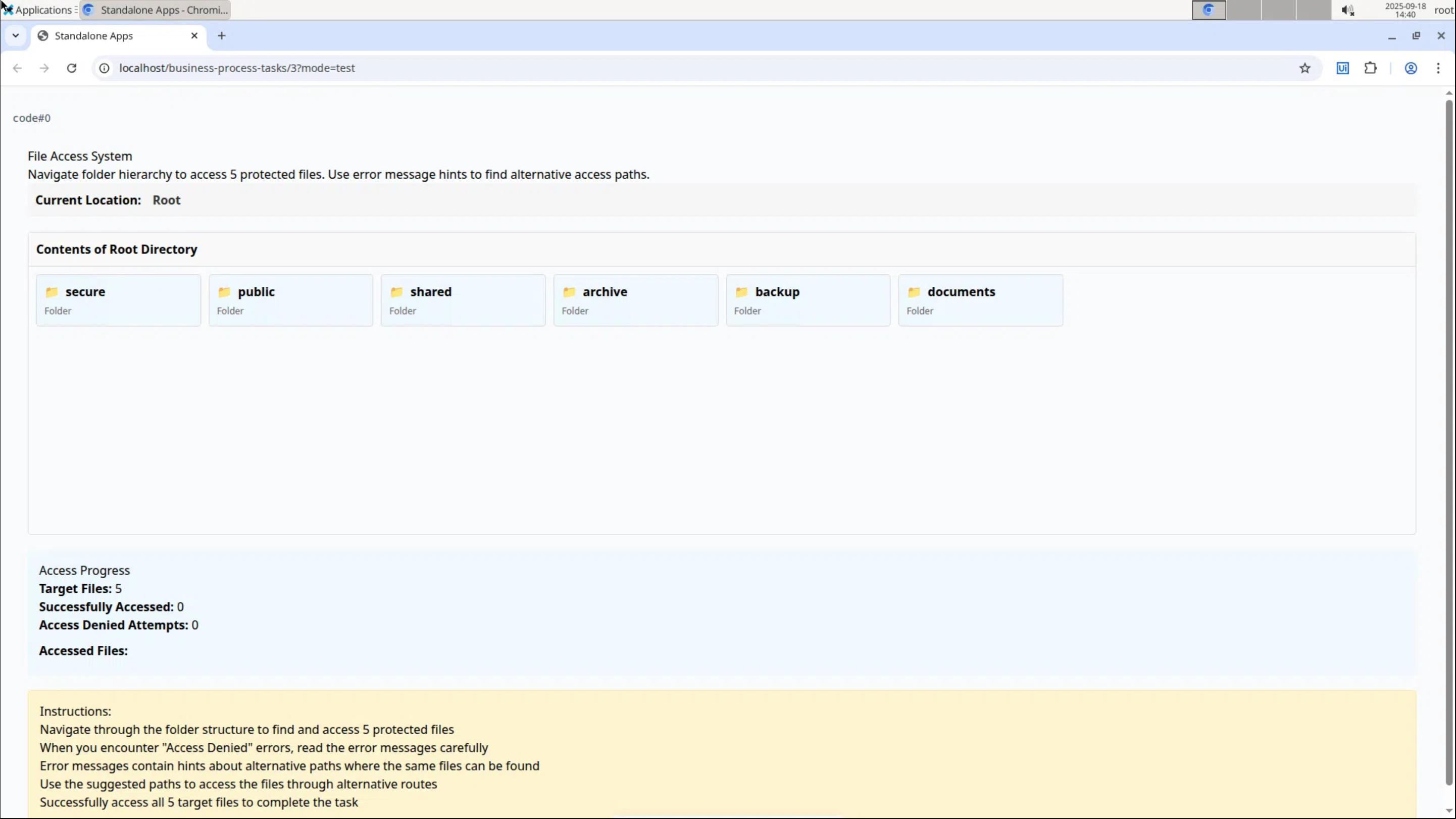}
    \caption{Agent gets stuck in a loop, repeatedly attempting the same failed action without exploration or recovery strategies.}
    \label{fig:stuck-in-loop-1}
\end{figure}

\begin{figure}[htbp]
    \centering
    \includegraphics[width=1.0\columnwidth]{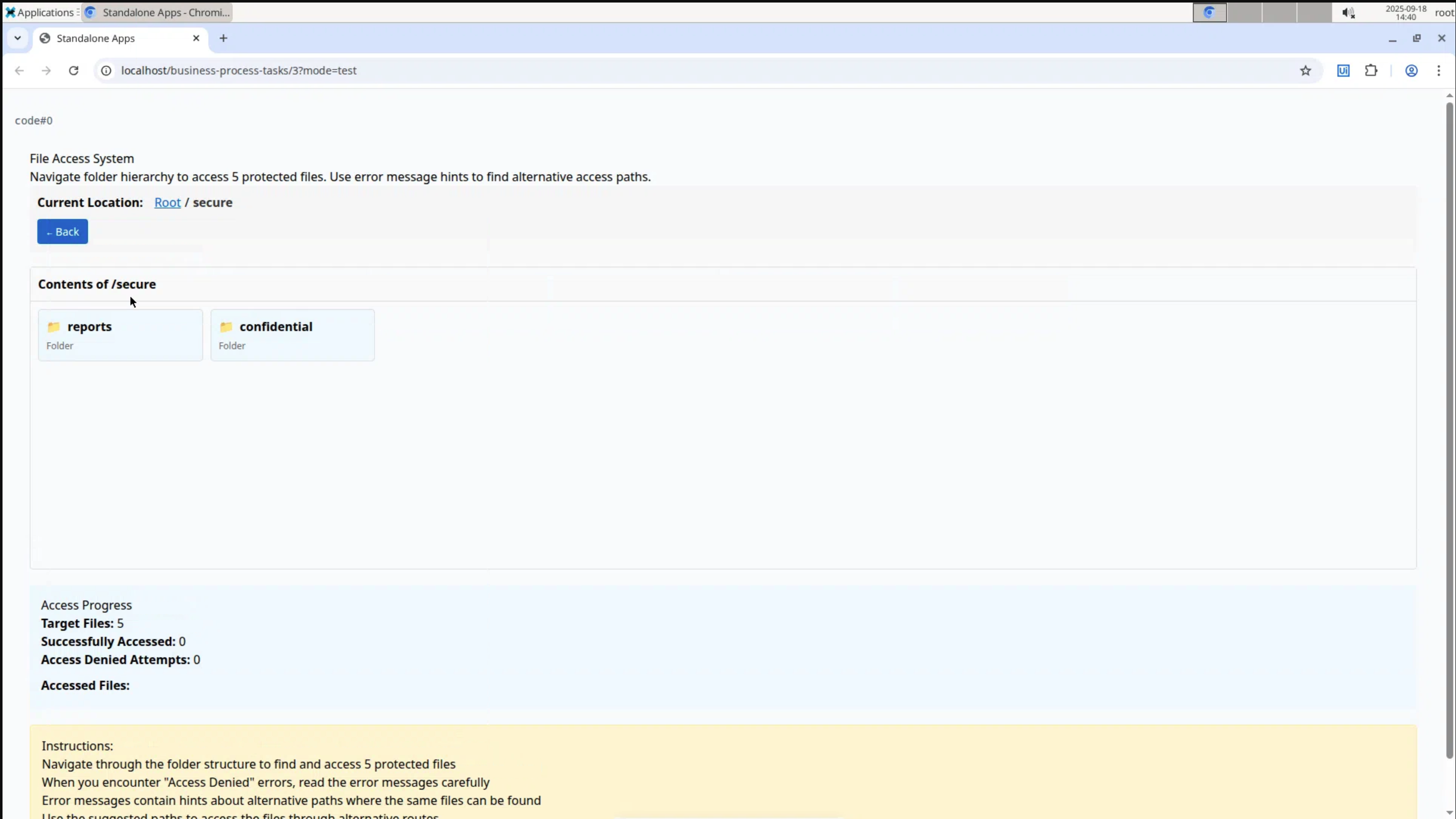}
    \caption{Agent gets stuck in a loop (continued).}
    \label{fig:stuck-in-loop-2}
\end{figure}

\begin{figure}[htbp]
    \centering
    \includegraphics[width=1.0\columnwidth]{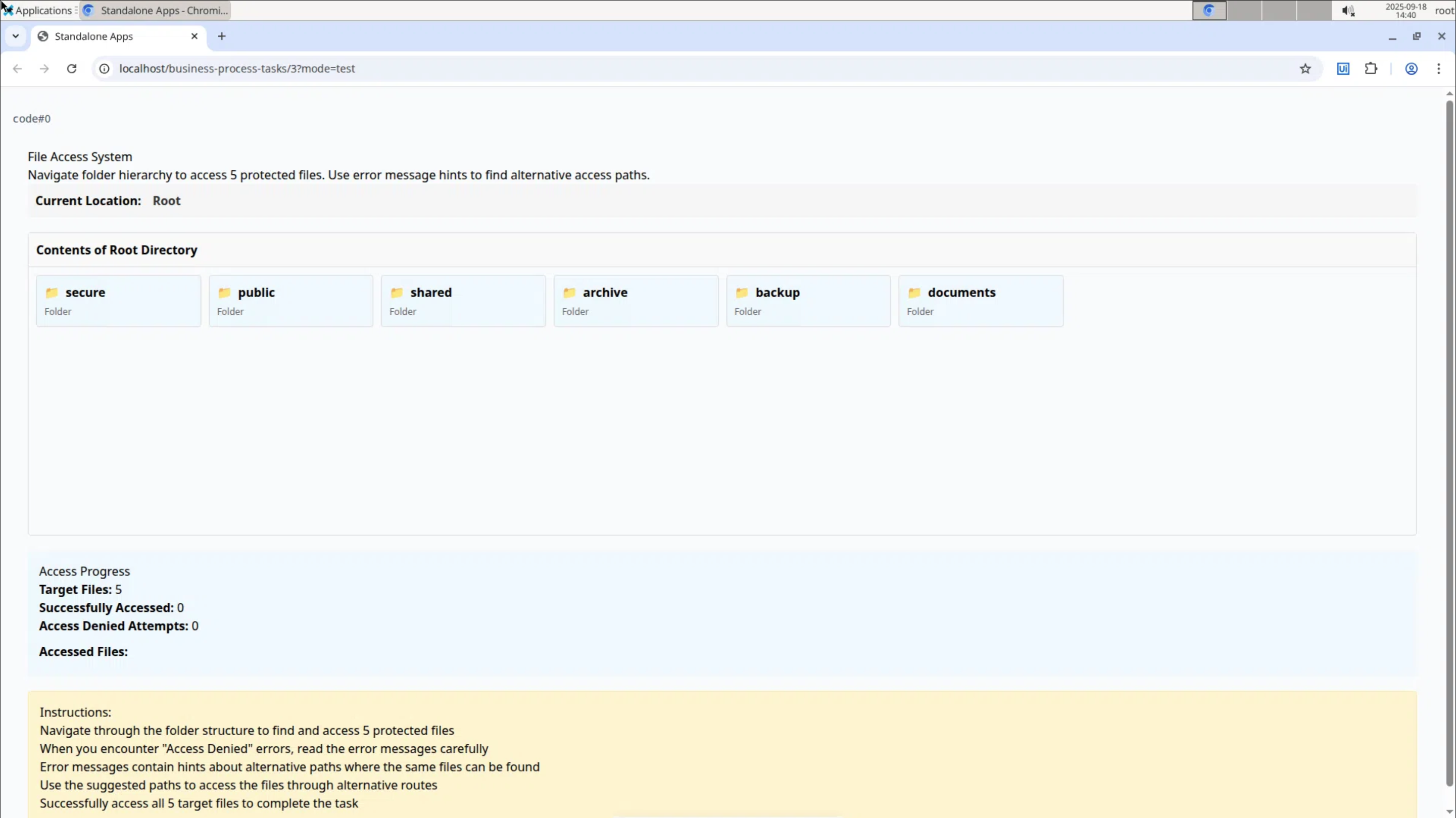}
    \caption{Agent gets stuck in a loop (final).}
    \label{fig:stuck-in-loop-3}
\end{figure}

Representative failure cases are illustrated in Figures~\ref{fig:misclick}, \ref{fig:incorrect-extraction}, and \ref{fig:stuck-in-loop-1}--\ref{fig:stuck-in-loop-3}.

\section{Conclusion}

We have introduced UI-CUBE, a benchmark that focuses on daily, repetitive tasks typically performed by human operators in organizations. This perspective highlights the gap between current AI capabilities and the practical needs of enterprise environments: while state-of-the-art models achieve 67-85\% success on simple UI interactions, they drop precipitously to 9-19\% on complex workflows. Even human evaluators---achieving near-perfect performance on simple tasks (97.9\%)---succeed only 61.2\% of the time on complex enterprise workflows when encountering unfamiliar applications.

Our evaluation shows that the latest models remain far from reliably solving complex tasks end to end, failing catastrophically at coordinated workflows requiring memory management, state tracking, and multi-step coordination. However, they perform very well on simple, atomic UI interactions. This creates an opportunity to use computer-use agents effectively by decomposing complex problems into workflows of simpler steps that current models can reliably handle.

Despite progress in code generation and general capabilities, current models still struggle with tasks that are straightforward for human operators, underscoring their limitations in real-world usage. Correctness remains a critical requirement in enterprise scenarios, but hallucinations continue to appear, often in subtle ways that are difficult to detect. This motivated UI-CUBE's execution-based validation approach: each task uses programmatic oracles that inspect final application state rather than relying on trajectory-based or LLM-as-judge evaluation, ensuring deterministic and reproducible assessment.

To make the benchmark feasible and reproducible, we implemented tasks in a controlled web environment with Docker containerization, which allowed us to mock enterprise applications, isolate behaviors, and construct deterministic scenarios with precise state reset capabilities. Importantly, the interactions we study are not specific to the web---they occur in desktop environments as well. The benchmark is designed to be easily extensible, and we plan to broaden it further to explicitly include desktop applications. UI-CUBE functions as an enterprise-readiness diagnostic, revealing that while current CUAs can interact with individual UI elements, they cannot yet function as reliable workflow automation tools.

\begin{acks}
We thank the anonymous reviewers for their valuable feedback.
\end{acks}



\begin{thebibliography}{13}

\bibitem{yao2022react}
Shunyu Yao, Jeffrey Zhao, Dian Yu, Nan Du, Izhak Shafran, Karthik Narasimhan, and Yuan Cao.
\newblock ReAct: Synergizing Reasoning and Acting in Language Models.
\newblock \emph{arXiv preprint arXiv:2210.03629}, 2022.

\bibitem{qin2025uitars}
Yujia Qin, Yining Ye, Junjie Fang, Haoming Wang, Shihao Liang, Shizuo Tian, Junda Zhang, Jiahao Li, Yunxin Li, Shijue Huang, Wanjun Zhong, Kuanye Li, Jiale Yang, Yu Miao, Woyu Lin, Longxiang Liu, Xu Jiang, Qianli Ma, Jingyu Li, Xiaojun Xiao, Kai Cai, Chuang Li, Yaowei Zheng, Chaolin Jin, Chen Li, Xiao Zhou, Minchao Wang, Haoli Chen, Zhaojian Li, Haihua Yang, Haifeng Liu, Feng Lin, Tao Peng, Xin Liu, and Guang Shi.
\newblock UI-TARS: Pioneering Automated GUI Interaction with Native Agents.
\newblock \emph{arXiv preprint arXiv:2501.12326}, 2025.

\bibitem{openai2023gpt4}
OpenAI.
\newblock GPT-4 Technical Report.
\newblock \emph{arXiv preprint arXiv:2303.08774}, 2023.

\bibitem{comanici2025gemini}
Gheorghe Comanici, Eric Bieber, Mike Schaekermann, Ice Pasupat, Noveen Sachdeva, and others.
\newblock Gemini 2.5: Pushing the Frontier with Advanced Reasoning, Multimodality, Long Context, and Next-Generation Agentic Capabilities.
\newblock \emph{arXiv preprint arXiv:2507.06261}, 2025.

\bibitem{shi2017world}
Tianlin Shi, Andrej Karpathy, Linxi Fan, Jonathan Hernandez, and Percy Liang.
\newblock World of Bits: An Open-Domain Platform for Web-Based Agents.
\newblock \emph{Proceedings of the 34th International Conference on Machine Learning (ICML)}, 2017.

\bibitem{liu2018reinforcement}
Evan Zheran Liu, Kelvin Guu, Panupong Pasupat, Tianlin Shi, and Percy Liang.
\newblock Reinforcement Learning on Web Interfaces Using Workflow-Guided Exploration.
\newblock \emph{International Conference on Learning Representations (ICLR)}, 2018.

\bibitem{deng2023mind2web}
Xiang Deng, Yu Gu, Boyuan Zheng, Shijie Chen, Samuel Stevens, Boshi Wang, Huan Sun, and Yu Su.
\newblock Mind2Web: Towards a Generalist Agent for the Web.
\newblock \emph{Advances in Neural Information Processing Systems (NeurIPS)}, 2023.

\bibitem{xie2024osworld}
Tianbao Xie, Danyang Zhang, Jixuan Chen, Xiaochuan Li, Siheng Zhao, Ruisheng Cao, Toh Jing Hua, Zhoujun Cheng, Dongchan Shin, Fangyu Lei, Yitao Liu, Yiheng Xu, Shuyan Zhou, Silvio Savarese, Caiming Xiong, Victor Zhong, and Tao Yu.
\newblock OSWorld: Benchmarking Multimodal Agents for Open-Ended Tasks in Real Computer Environments.
\newblock \emph{arXiv preprint arXiv:2404.07972}, 2024.

\bibitem{abhyankar2025osworld}
Reyna Abhyankar, Qi Qi, and Yiying Zhang.
\newblock OSWorld-Human: Benchmarking the Efficiency of Computer-Use Agents.
\newblock \emph{arXiv preprint arXiv:2506.16042}, 2025.

\bibitem{zhou2023webarena}
Shuyan Zhou, Frank F. Xu, Hao Zhu, Xuhui Zhou, Robert Lo, Abishek Sridhar, Xianyi Cheng, Yonatan Bisk, Daniel Fried, Uri Alon, Tianyue Ou, and Graham Neubig.
\newblock WebArena: A Realistic Web Environment for Building Autonomous Agents.
\newblock \emph{arXiv preprint arXiv:2307.13854}, 2023.

\bibitem{jimenez2023swe}
Carlos E. Jimenez, John Yang, Alexander Wettig, Shunyu Yao, Kexin Pei, Ofir Press, and Karthik Narasimhan.
\newblock SWE-bench: Can Language Models Resolve Real-World GitHub Issues?
\newblock \emph{arXiv preprint arXiv:2310.06770}, 2023.

\bibitem{xu2024theagentcompany}
Frank F. Xu, Yufan Song, Boxuan Li, Yuxuan Tang, Kritanjali Jain, Mengxue Bao, Zora Z. Wang, Xuhui Zhou, Zhitong Guo, Murong Cao, Mingyang Yang, Hao Yang Lu, Amaad Martin, Zhe Su, Leander Maben, Raj Mehta, Wayne Chi, Lawrence Jang, Yiqing Xie, Shuyan Zhou, Graham Neubig, and Daniel Fried.
\newblock TheAgentCompany: Benchmarking LLM Agents on Consequential Real-World Tasks.
\newblock \emph{arXiv preprint arXiv:2412.14161}, 2024.

\bibitem{he2024webvoyager}
Hongliang He, Wenlin Yao, Kaixin Ma, Wenhao Yu, Yong Dai, Hongming Zhang, Zhenzhong Lan, and Dong Yu.
\newblock WebVoyager: Building an End-to-End Web Agent with Large Multimodal Models.
\newblock \emph{arXiv preprint arXiv:2401.13919}, 2024.

\bibitem{dai2025scuba}
Yutong Dai, Krithika Ramakrishnan, Jing Gu, Matthew Fernandez, Yanqi Luo, Viraj Prabhu, Zhenyu Hu, Silvio Savarese, Caiming Xiong, Zeyuan Chen, and Ran Xu.
\newblock SCUBA: Salesforce Computer Use Benchmark.
\newblock \emph{arXiv preprint arXiv:2509.26506}, 2025.

\bibitem{boisvert2024workarena}
Léo Boisvert, Megh Thakkar, Maxime Gasse, Massimo Caccia, Thibault Le Sellier De Chezelles, Quentin Cappart, Nicolas Chapados, Alexandre Lacoste, and Alexandre Drouin.
\newblock WorkArena++: Towards Compositional Planning and Reasoning-based Common Knowledge Work Tasks.
\newblock \emph{arXiv preprint arXiv:2407.05291}, 2024.

\bibitem{huang2025crmarenapro}
Kung-Hsiang Huang, Akshara Prabhakar, Onkar Thorat, Divyansh Agarwal, Prafulla Kumar Choubey, Yixin Mao, Silvio Savarese, Caiming Xiong, and Chien-Sheng Wu.
\newblock CRMArena-Pro: Holistic Assessment of LLM Agents Across Diverse Business Scenarios and Interactions.
\newblock \emph{arXiv preprint arXiv:2505.18878}, 2025.

\bibitem{garg2025real}
Tanmay Garg, Junayed Mahmud, Jayati Deshmukh, and Sriram Gopalakrishnan.
\newblock REAL: Benchmarking Autonomous Agents on Deterministic Simulations of Real Websites.
\newblock \emph{arXiv preprint arXiv:2504.11543}, 2025.

\bibitem{wang2024officebench}
Zilong Wang, Yuedong Cui, Li Zhong, Zimin Zhang, Guiyang Hou, Taicheng Guo, Xuanming Zhang, Yijie Wang, Shengze Xu, and Lu Chen.
\newblock OfficeBench: Benchmarking Language Agents across Multiple Applications for Office Automation.
\newblock \emph{arXiv preprint arXiv:2407.19056}, 2024.

\bibitem{wang2024agentmemory}
Zora Zhiruo Wang, Jiayuan Mao, Daniel Fried, and Graham Neubig.
\newblock Agent Workflow Memory.
\newblock \emph{arXiv preprint arXiv:2409.07429}, 2024.

\bibitem{wang2025mmbench}
Junyang Wang, Haiyang Xu, Jiabo Ye, Hongze Liu, Guohai Xu, Yiyang Zhou, Chengyi Wang, Tong Lu, and Ming Yan.
\newblock MMBench-GUI: Hierarchical Multi-Platform Evaluation Framework for GUI Agents.
\newblock \emph{arXiv preprint arXiv:2507.19478}, 2025.

\bibitem{bonatti2024windows}
Rogerio Bonatti, Dan Zhao, Francesco Bonacci, Dillon Dupont, Sara Abdali, Yinheng Li, Justin Wagle, Kazuhito Koishida, Arthur Bucker, Lawrence Jang, and Shital Shah.
\newblock Windows Agent Arena: Evaluating Multi-Modal OS Agents at Scale.
\newblock \emph{arXiv preprint arXiv:2409.08264}, 2024.

\bibitem{rawles2024androidworld}
Christopher Rawles, Sarah Clinckemaillie, Yifan Chang, Jonathan Waltz, Gabrielle Lau, Marybeth Fair, Alice Li, William Bishop, Wei Li, Folawiyo Campbell-Ajala, Daniel Toyama, Robert Berry, Divya Tyamagundlu, Timothy Lillicrap, and Oriana Riva.
\newblock AndroidWorld: A Dynamic Benchmarking Environment for Autonomous Agents.
\newblock \emph{arXiv preprint arXiv:2405.14573}, 2024.

\bibitem{yang2025macosworld}
Zitong Yang, Jun Li, Siyuan Hu, Wanqing Xie, Shulin Li, Xinyu Guan, Tian He, Kun Zhou, and Wayne Xin Zhao.
\newblock macOSWorld: A Multilingual Interactive Benchmark for GUI Agents.
\newblock \emph{arXiv preprint arXiv:2506.04135}, 2025.

\end{thebibliography}
\end{document}